\documentclass[aps,nofootinbib,floatfix]{revtex4}

\usepackage{amssymb}
\usepackage{amsmath}
\usepackage{graphicx}
\begin{document}

\title{Holographic Dark Energy Model with Hubble Horizon as an IR Cut-off}

\author{Lixin Xu\footnote{Corresponding author}}
\email{lxxu@dlut.edu.cn}

\affiliation{Institute of Theoretical Physics, School of Physics \&
Optoelectronic Technology, Dalian University of Technology, Dalian,
116024, P. R. China}

\begin{abstract}
The main task of this paper is to realize a cosmic observational
compatible universe in the framework of holographic dark energy
model when the Hubble horizon $H$ is taken as the role of an IR
cut-off. When the model parameter $c$ of a time variable
cosmological constant (CC) $\Lambda(t)=3c^{2}H^{2}(t)$ becomes time
or scale dependent, an extra term enters in the effective equation
of sate (EoS) of the vacuum energy $w^{eff}_{\Lambda}=-c^2-d\ln
c^{2}/3d\ln a$. This extra term can make the effective EoS of time
variable CC cross the cosmological boundary and be phantom-like at
present. For the lack of a first principle and fundamental physics
theory to obtain the form $c^2$, we give a simple parameterized form
of $c^2$ as an example. Then the model is confronted by the cosmic
observations including SN Ia, BAO and CMB shift parameter $R$. The
result shows that the model is consistent with cosmic observations.
\end{abstract}

%\pacs{Added}

\keywords{time variable cosmological constant; dark energy} \hfill
ITP-DUT/2009-07

\maketitle

\section{Introduction}

The observation of the Supernovae of type Ia
\cite{ref:Riess98,ref:Perlmuter99} provides the evidence that the
universe is undergoing accelerated expansion at present. Combining
the observations from Cosmic Background Radiation
\cite{ref:Spergel03,ref:Spergel06} and SDSS
\cite{ref:Tegmark1,ref:Tegmark2}, one concludes that the universe at
present is dominated by $70\%$ exotic component, dubbed dark energy,
which has negative pressure and pushes the universe to accelerated
expansion. Of course, a natural explanation to the accelerated
expansion is due to a positive tiny cosmological constant. Though,
it suffers the so-called {\it fine tuning} and {\it cosmic
coincidence} problems. However, in $2\sigma$ confidence level, it
fits the observations very well \cite{ref:Komatsu}. If the
cosmological constant is not a real constant but is time variable,
the fine tuning and cosmic coincidence problems can be removed. In
fact, this possibility was considered in the past years.

In particular, the dynamic vacuum energy density based on
holographic principle was investigated extensively
\cite{ref:holo1,ref:holo2}. According to the holographic principle,
the number of degrees of freedom in a bounded system should be
finite and has relations with the area of its boundary. By applying
the principle to cosmology, one can obtain the upper bound of the
entropy contained in the universe. For a system with size $L$ and UV
cut-off $\Lambda$ without decaying into a black hole, it is required
that the total energy in a region of size $L$ should not exceed the
mass of a black hole of the same size, thus $L^3\rho_{\Lambda} \le L
M^2_{P}$. The largest $L$ allowed is the one saturating this
inequality, thus $\rho_{\Lambda} =3c^2 M^{2}_{P} L^{-2}$, where $c$
is a numerical constant and $M_{P}$ is the reduced Planck Mass
$M^{-2}_{P}=8 \pi G$. It just means a {\it duality} between UV
cut-off and IR cut-off. The UV cut-off is related to the vacuum
energy, and IR cut-off is related to the large scale of the
universe, for example Hubble horizon, future event horizon or
particle horizon which were discussed by
\cite{ref:holo1,ref:holo2,ref:Horvat1,ref:Horvat2}. The holographic
dark energy in Brans-Dicke theory was also studied in Ref.
\cite{ref:BransDicke,ref:BDH1,ref:BDH2,ref:BDH3,ref:BDH4,ref:BDH5}.
In the standard and Brans-Dicke holographic dark energy models when
the Hubble horizon is taken as the role of IR cut-off,
non-accelerated expansion universe can be achieved
\cite{ref:holo1,ref:holo2,ref:BransDicke}. However, the Hubble
horizon is the most natural cosmological length scale, how to
realize an accelerated expansion by taking it as an IR cut-off will
be interesting.

Furthermore, the {\it holographic cosmological constant} were
discussed in \cite{ref:Horvat1,ref:Horvat2,ref:Feng}, where a time
variable cosmological constant comes from the holographic principle.
Inspired by the observation of the relation between cosmological
length or time scale with any nonzero value of the cosmological
constant $r_{\Lambda}=t_{\Lambda}=\sqrt{3/|\Lambda|}$, {\it horizon
cosmological constants} were discussed in \cite{Xu:TVCC}. In these
two cases, an accelerated expansion universe could be obtained at
present, precisely speaking a scaling solution was obtained, when
the Hubble horizon was taken as the role of an IR cut-off. But
unfortunately, non-transition from decelerated expansion to
accelerated expansion can be realized in this scenario. This
observation motivates us to consider the possibility of realizing
accelerated expansion by mini modification of holographic or horizon
cosmological constant model. This will be the main task of this
work.

This paper is structured as follows. In Section \ref{sec:TVC}, we
give a brief review of time variable cosmological constant. In
Section \ref{sec:HIR}, Hubble horizon as an IR cut-off will be
explored when $c$ is fixed constant and time or scale dependent
respectively. In this section, cosmic observational constraint is
also implemented. Where the cosmic observations and constraint
methods are put in the Appendix \ref{app:constraints}. Conclusions
are set in Section \ref{sec:Con}.

\section{Time Variable Cosmological Constant}\label{sec:TVC}

The Einstein equation with a cosmological constant is written as
\begin{equation}
R_{\mu\nu}-\frac{1}{2}Rg_{\mu\nu}+\Lambda g_{\mu\nu}=8\pi G
T_{\mu\nu},\label{eq:EE}
\end{equation}
where $T_{\mu\nu}$ is the energy-momentum tensor of ordinary matter
and radiation. From the Bianchi identity, one has the conservation
of the energy-momentum tensor $\nabla^{\mu}T_{\mu\nu}=0$, it follows
necessarily that $\Lambda$ is a constant. To have a time variable
cosmological constant $\Lambda=\Lambda(t)$, one can move the
cosmological constant to the right hand side of Eq. (\ref{eq:EE})
and take $\tilde{T}_{\mu\nu}=T_{\mu\nu}-\frac{\Lambda(t)}{8\pi
G}g_{\mu\nu}$ as the total energy-momentum tenor. Once again to
preserve the Bianchi identity or local energy-momentum conservation
law, $\nabla^{\mu}\tilde{T}_{\mu\nu}=0$, one has, in a spatially
flat FRW universe,
\begin{equation}
\dot{\rho}_{\Lambda}+\dot{\rho}_{m}+3H\left(1+w_{m}\right)\rho_{m}=0,\label{eq:conservation}
\end{equation}
where $\rho_{\Lambda}=M^2_{P}\Lambda(t)$ is the energy density of
time variable cosmological constant and its equation of state is
$w_{\Lambda}=-1$, and $w_{m}$ is the equation of state of ordinary
matter, for dark matter $w_m=0$. It is natural to consider
interactions between variable cosmological constant and dark matter
\cite{ref:Horvat2}, as seen from Eq. (\ref{eq:conservation}). After
introducing an interaction term $Q$, one has
\begin{eqnarray}
\dot{\rho}_{m}+3H\left(1+w_{m}\right)\rho_{m}=Q,\label{eq:rhom} \\
\dot{\rho}_{\Lambda}+3H\left(\rho_{\Lambda}+p_{\Lambda}\right)=-Q,\label{eq:rholambda}
\end{eqnarray}
and the total energy-momentum conservation equation
\begin{equation}
\dot{\rho}_{tot}+3H\left(\rho_{tot}+p_{tot}\right)=0.
\end{equation}
For a time variable cosmological constant, the equality
$\rho_{\Lambda}+p_{\Lambda}=0$ still holds. Immediately, one has the
interaction term $Q=-\dot{\rho}_{\Lambda}$ which is different from
the interactions between dark matter and dark energy considered in
the literatures \cite{ref:interaction} where a general interacting
form $Q=3b^2H\left(\rho_{m}+\rho_{\Lambda}\right)$ is put by hand.
With observation to Eq. (\ref{eq:rholambda}), the interaction term
$Q$ can be moved to the left hand side of the equation, and one has
the effective pressure of the time variable cosmological constant-
dark energy
\begin{equation}
\dot{\rho}_{\Lambda}+3H\left(\rho_{\Lambda}+p^{eff}_{\Lambda}\right)=0,
\end{equation}
where $p^{eff}_{\Lambda}=p_{\Lambda}+\frac{Q}{3H}$ is the effective
dark energy pressure. Also, one can define the effective equation of
state of dark energy
\begin{eqnarray}
w^{eff}_{\Lambda}&=&\frac{p^{eff}_{\Lambda}}{\rho_{\Lambda}}\nonumber\\
&=&-1+\frac{Q}{3H\rho_{\Lambda}}\nonumber\\
&=&=-1-\frac{1}{3}\frac{d \ln \rho_{\Lambda}}{d\ln
a}.\label{eq:EEOS}
\end{eqnarray}
The Friedmann equation as usual can be written as, in a spatially
flat FRW universe,
\begin{equation}
H^2=\frac{1}{3M^2_P}\left(\rho_{m}+\rho_{\Lambda}\right)\label{eq:FE}.
\end{equation}

\section{Hubble horizon as an IR cut-off}\label{sec:HIR}

\subsection{Fixed constant $c$}

Horvat has considered a time variable cosmological constant from
holographic principle \cite{ref:Horvat2}, where the Hubble horizon
$H^{-1}$ was taken as a cosmological length scale. The time variable
cosmological constant is given by \cite{ref:Horvat2}
\begin{equation}
\Lambda(t)=3c^2H^{2}(t),
\end{equation}
where $c$ is a fixed constant. As known, our universe is filled with
dark matter and dark energy and deviates from a de Sitter one. Just
to describe this gap, the constant $c$ was introduced. With this
observation, $c$ can be named gap filling parameter. It can be seen
that a $c^2<1$ constant is expected under the consideration of the
energy budget of the universe. Also, one can see that a de Sitter
universe will be recovered when $c^2=1$ is respected. Now, the
corresponding vacuum energy density can be written as
\begin{equation}
\rho_{\Lambda}=3c^2M^2_{P}H^2
\end{equation}
which takes the same form as the so-called holographic dark energy
based on holographic principle. With this vacuum energy, the
Friedmann equation (\ref{eq:FE}) can be rewritten as
\begin{equation}
\rho_{m}=3(1-c^2)M^2_PH^2.
\end{equation}
To protect a positive dark matter energy density, a constraint
\begin{equation}
c^2<1
\end{equation}
is required. Immediately, a scaling solution is
obtained
\begin{equation}
\frac{\rho_{m}}{\rho_{\Lambda}}=\frac{1-c^2}{c^2}\label{eq:scaling}.
\end{equation}
Substituting Eq. (\ref{eq:scaling}) into Eq.
(\ref{eq:conservation}), one has
\begin{equation}
\rho_{\Lambda}=\frac{c^2}{1-c^2}\rho_{m}\sim a^{-3(1-c^2)}.
\end{equation}
Here, one can see a rather different result on $\rho_{m}$ from the
standard evolution $a^{-3}$. In this case, the deceleration
parameter becomes
\begin{eqnarray}
q&=&-\frac{\ddot{a}a}{\dot{a}^2}=-\frac{\dot{H}+H^2}{H^2}\nonumber\\
&=&\frac{1}{2}-\frac{3}{2}c^2.
\end{eqnarray}
To obtain a current accelerated expansion universe, i.e. $q<0$, and
to protect positivity of dark matter energy density, one obtains a
constraint to the constant $c$
\begin{equation}
1/3<c^2<1.\label{eq:cconstraint}
\end{equation}
The effective equation of state of vacuum energy density is
\begin{eqnarray}
w^{eff}_{\Lambda}&=&-1-\frac{1}{3}\frac{d \ln \rho_{\Lambda}}{d\ln
a}\nonumber\\
&=&-c^2.
\end{eqnarray}
Under the constraint Eq.(\ref{eq:cconstraint}), one can see that a
quintessence like dark energy is obtained. This is tremendous
different from holographic dark energy model where non-accelerated
expansion universe can be achieved when the Hubble horizon taken as
the role of an IR cut-off \cite{ref:holo1,ref:holo2,ref:BransDicke}.
Also, it is easily see that the de Sitter universe will be recovered
when $c^2=1$ is respected. Once the constant $c^2$ deviates from
$c^2=1$, a scaling solution will be obtained.

\subsection{Time Variable constant c}

It is clear from the above subsection that when $c$ is a fixed
constant, non-transition from decelerated expansion to accelerated
expansion can be realized. And, a possible remedy maybe make the
constant $c$ not fixed but time or scale dependent. A time variable
$c$ was considered in \cite{ref:Zimdahl} to solve the coincidence
problem. So, we assume that $c$ is time variable or scale dependent,
i.e,
\begin{equation}
\rho_{\Lambda}=3c^2(t)M^2_{P}H^2.
\end{equation}
As that of $c$ a fixed constant case, one also has the relation
\begin{equation}
\rho_{m}=3M^{2}_{P}(1-c^{2}(t))H^{2}.
\end{equation}
Also, to protect energy density of cold dark matter from negativity, the constraint $c^{2}<1$ is required.
From the conservation equation of cold dark matter Eq. (\ref{eq:rhom}) and the Friedmann equation, one has
\begin{equation}
(1+z)\frac{d\ln H}{dz}-\frac{3}{2}(1-c^{2}(z))=0\label{eq:hz}.
\end{equation}
To solve the Eq. (\ref{eq:hz}), one has to assume some concrete
forms of the parameter $c(z)$. After simple calculation, one also
has the same form of the deceleration parameter as the case of the
fixed constant $c$
\begin{equation}
q=\frac{1}{2}-\frac{3}{2}c^2(z).
\end{equation}
One can easily find that once $0<c^{2}(z)<1$ is time or scale
dependent, the possible transition from deceleration expansion to
accelerated expansion can be realized. However, one will derive a
different form of effective EoS of the time variable CC in the case
of time or scale dependence of parameter $c$
\begin{eqnarray}
w^{eff}_{\Lambda}&=&-1-\frac{1}{3}\frac{d \ln \rho_{\Lambda}}{d\ln
a}\nonumber\\
&=&-c^2-\frac{1}{3}\frac{d\ln c^{2}}{d\ln a}.
\end{eqnarray}
Here, an extra term enters in the effective EoS and can make the EoS
cross the CC boundary and be phantom-like at present. Also, by the
definition of dimensionless energy density of time variable CC
$\Omega_{\Lambda}=\rho_{\Lambda}/(3M^{2}_{P}H^{2})$, one obtains the
simple form
\begin{equation}
\Omega_{\Lambda}=c^{2}(t).
\end{equation}
Obviously, it is time or scale dependent as a contrast to the fixed
constant $c$ case.

The next step is to give some forms of time or scale dependent
parameter $c^{2}$. However, unfortunately we have no any first
principle and underlying physics theory to obtain the forms of
$c^{2}$ at present. We only know that the constraint
$1/3<c^{2}(z=0)<1$ must be satisfied to have an accelerated
expansion universe at present. Also, the transition from decelerated
expansion to accelerated expansion would also be covered
potentially. And, the tension of parameters contained in the
parameterized form of $c^{2}$ must be as looser as possible. In
fact, we can reverse the process by giving some parameterized forms
of the deceleration parameter. For example, we can assume the form
of deceleration parameter in redshift $z$ as follows
\begin{equation}
q(z)=q_{0}+q_{1}\frac{z}{1+z},
\end{equation}
which has been discussed in \cite{ref:XU}. Then, one immediately has
the parameterized form of $c^{2}$
\begin{equation}
c^{2}(z)=\frac{1}{3}(1-2q_{0})-\frac{2q_{1}}{3}\frac{z}{1+z}.
\end{equation}
As required the condition $c^{2}(z)\rightarrow 0$ would be satisfied
at early epoch, when $z\rightarrow \infty$. One has the relation
between $q_{0}$ and $q_{1}$
\begin{equation}
q_{0}+q_{1}=\frac{1}{2}.
\end{equation}
Then, $c^{2}(z)$ can be rewritten as
\begin{equation}
c^{2}(z)=\frac{1}{3}(1-2q_{0})\frac{1}{1+z}.
\end{equation}
Taken this parameterization as a clue, an generalized form of
$c^2(z)$ can be assumed as the form of
\begin{equation}
c^{2}(z)=\frac{a}{(1+z)^{b}},
\end{equation}
where $\Omega_{\Lambda 0}=a\ge 0$ and $b\ge 0$ are model parameters
which can be determined by cosmic observations. It is clear that our
model is a one parameter model. Also, one can easily has the
expression of the deceleration parameter
\begin{equation}
q(z)=\frac{1}{2}-\frac{3}{2}\frac{a}{(1+z)^{b}}
\end{equation}
Now, the Eq. (\ref{eq:hz}) can be integrated and the solution is
\begin{equation}
H(z)=H_{0}(1+z)^{3/2}\exp\left\{\frac{3a}{2b}\left[(1+z)^{-b}-1\right]\right\}.
\end{equation}
Having this form of Hubble parameter, the model can be confronted by
cosmic observations, such as SN Ia, BAO and CMB shift parameter $R$.
In this paper, the (SCP) Union sample including $307$ SN, ration
$D_{V}(0.35)/D_{V}(0.2)$ detected by BAO and CMB shift parameter $R$
from the WMAP5 are used, for the details please see the Appendix
\ref{app:constraints}. The likelihood function is given by $L\propto
e^{-\chi^2/2}$, where $\chi^2$ is
\begin{equation}
\chi^2=\chi^2_{SNIa}+\chi^2_{BAO}+\chi^2_{CMB},
\end{equation}
$\chi^2_{SN}$ is given in Eq. (\ref{eq:chi2SN}), $\chi^2_{BAO}$ is
given in Eq. (\ref{eq:chi2BAO}), $\chi^2_{CMB}$ is given in Eq.
(\ref{eq:chi2CMB}).
%In this paper, the central values
%of $\Omega_b h^2=0.02265\pm 0.00059$, $\Omega_m h^2=0.1369\pm
%0.0037$ from 5-year {\it WMAP} results \cite{ref:Komatsu2008} and
%$H_0=72\pm{\rm 8 km s^{-1}Mpc^{-1}}$ are adopted.
After calculation,
the results are listed in Tab. \ref{tab:result}.
\begin{table}[tbh]
\begin{center}
\begin{tabular}{c|c|c|c|c}
\hline\hline Datasets & $\chi^2_{min}$ & $a=\Omega_{\Lambda0}(1\sigma)$ & $b (1\sigma)$ & $z_{T}(1\sigma)$\\
\hline SN+BAO+CMB & $313.261$  & $0.764^{+0.012}_{-0.013}$ & $1.480^{+0.054}_{-0.050}$ & $0.751^{+0.122}_{-0.108}$\\
\hline
\end{tabular}
\caption{The minimum values of $\chi^2$ and best fit values of the
parameters.}\label{tab:result}
\end{center}
\end{table}

\begin{figure}[tbh]
\centering
\includegraphics[width=5.0in]{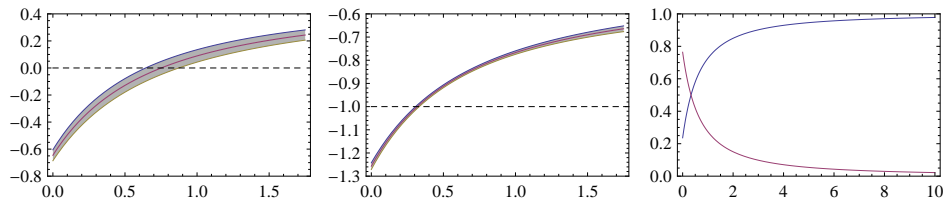}
\caption{The evolution curves of $q(z)$ (left panel),
$w^{eff}_{\Lambda}(z)$ (central panel) with $1\sigma$ error region
and dimensionless parameters $\Omega_{m}(z)$ and
$\Omega_{\Lambda}(z)$ (right panel) with respect to redshift $z$
where the best fit values are adopted.}\label{fig:qwomega}
\end{figure}

\begin{figure}[tbh]
\centering
\includegraphics[width=2.0in]{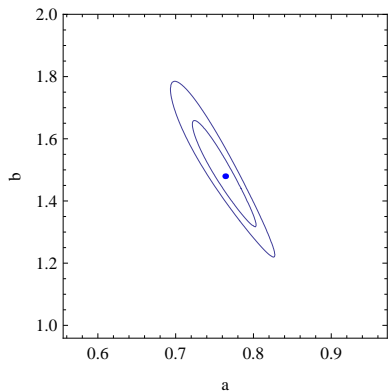}
\caption{The contours in the planes of $a-b$ with $1\sigma$ and
$2\sigma$ regions. The central dots denote the best fit values of
model parameters.}\label{fig:cons}
\end{figure}

With the best fit values of model parameters, the evolutions of
deceleration parameter, effective EoS of time variable CC and
dimensionless energy densities of time variable CC and cold dark
matter with respect to the redshift $z$ are plotted in Fig.
\ref{fig:qwomega}. Also the model parameter contours are plotted in
Fig. \ref{fig:cons}.
Clearly, with this simple parameterized form of $c^2$, an
observational consistent model is presented when the Hubble horizon
is taken as the role of an IR cut-off in the holographic dark energy
scenario. For the introduction of an extra term in the effective EoS
of the vacuum energy density of time variable cosmological constant,
the cosmological constant boundary crossing can be realized, as seen
in the central panel of Fig. \ref{fig:qwomega}. One can also see
that the effective EoS of time variable CC is phantom-like at
present.

\section{Conclusions}\label{sec:Con}

In this paper, time variable CC is explored when the Hubble horizon
is taken as the role of an IR cut-off, i.e.
$\Lambda(t)=3c^{2}H^{2}(t)$ which corresponds to the vacuum energy
density $\rho_{\Lambda}=3c^{2}M^{2}_{P}H^{2}$. When $c$ is a fixed
constant, a scaling solution is obtained. If $c$ is in the range of
$1/3<c^{2}<1$, an accelerated expansion universe can exist. But,
unfortunately with this fixed gap filling constant $c$,
no-transition from decelerated expansion to accelerated expansion
can be realized. However, the Hubble horizon is a natural choice of
cosmological length scale. To realize an accelerated expansion
universe, a transition from the past decelerated expansion to recent
accelerated expansion and cosmic observational compatible model in
the case of Hubble horizon as an IR cut-off, a time or scale
dependent gap filling constant $c$ is considered. With this time or
scale dependent $c$, a time or scale dependent dimensionless energy
density is derived. And, the effective EoS of time variable CC gains
an term which can make it cross cosmological constant boundary and
be phantom-like at present. By giving a simple parameterized form of
$c^2$ as an example, the model was confronted by cosmic observations
which include SN Ia, BAO and CMB shift parameter $R$. The constraint
result shows that a cosmic observational compatible model can be
realized in this framework when the Hubble horizon is taken as the
role of an IR cut-off. That can be seen from the Fig.
\ref{fig:qwomega}. However, we do not know the first principle or
fundamental physics theory to give the form of time or scale
dependent $c(a)$. It seems the limitation of our model. But, we
expect this consideration can shed light on the study of holographic
dark energy models.

\acknowledgements{L. Xu thanks Prof. Z. H. Zhu for his hospitality
during the author's visit in Beijing Normal University. This work is
supported by NSF (10703001), SRFDP (20070141034) of P.R. China.}

\appendix

\section{Cosmic Observational Constraints}\label{app:constraints}

In this section, cosmic observations and methods used in this paper
are described.

\subsection{SN Ia}
We constrain the parameters with the Supernovae Cosmology Project
(SCP) Union sample including $307$ SN Ia \cite{ref:SCP}, which is
distributed over the redshift interval $0.015\le z\le 1.551$.
Constraints from SN Ia can be obtained by fitting the distance
modulus $\mu(z)$
\cite{ref:smallomega,ref:BDSN1,ref:BDSN2,ref:BDSN3,ref:JCAPXU,ref:SNchi2}
\begin{equation}
\mu_{th}(z)=5\log_{10}(D_{L}(z))+\mu_{0},
\end{equation}
where, $D_{L}(z)$ is the Hubble free luminosity distance $H_0
d_L(z)/c$ and
\begin{eqnarray}
d_L(z)&=&c(1+z)\int_{0}^{z}\frac{dz^{\prime}}{H(z^{\prime})}\\
\mu_0&\equiv&42.38-5\log_{10}h,
\end{eqnarray}
where $H_0$ is the Hubble constant which is written in terms of a
re-normalized quantity $h$ defined as $H_0 =100 h~{\rm km ~s}^{-1}
{\rm Mpc}^{-1}$. The observed distance moduli $\mu_{obs}(z_i)$ of SN
Ia at $z_i$ is
\begin{equation}
\mu_{obs}(z_i) = m_{obs}(z_i)-M,
\end{equation}
where $M$ is their absolute magnitudes.

For the SN Ia dataset, the best fit values of parameters $p_s$  in
the model can be determined by a likelihood analysis based on the
calculation of
\begin{eqnarray}
\chi^2(p_s,M^{\prime})&\equiv& \sum_{SN}\frac{\left[
\mu_{obs}(z_i)-\mu_{th}(p_s,z_i)\right]^2} {\sigma_i^2} \nonumber\\
&=&\sum_{SN}\frac{\left[ 5 \log_{10}(D_L(p_s,z_i)) - m_{obs}(z_i) +
M^{\prime} \right]^2} {\sigma_i^2}, \label{eq:chi2}
\end{eqnarray}
where $M^{\prime}\equiv\mu_0+M$ is a nuisance parameter which
includes the absolute magnitude and $h$. The nuisance parameter
$M^{\prime}$ can be marginalized over analytically
\cite{ref:SNchi2},
\begin{equation}
\bar{\chi}^2(p_s) = -2 \ln \int_{-\infty}^{+\infty}\exp \left[
-\frac{1}{2} \chi^2(p_s,M^{\prime}) \right] dM^{\prime},
\label{eq:chi2marg}
\end{equation}
to arrive at
\begin{equation}
\bar{\chi}^2 =  A - \frac{B^2}{C} + \ln \left( \frac{C}{2\pi}\right)
, \label{eq:chi2mar}
\end{equation}
where
\begin{equation}
A=\sum_{SN} \frac {\left[5\log_{10}
(D_L(p_s,z_i))-m_{obs}(z_i)\right]^2}{\sigma_i^2},
\end{equation}
\begin{equation}
B=\sum_{SN} \frac {5
\log_{10}(D_L(p_s,z_i)-m_{obs}(z_i)}{\sigma_i^2},
\end{equation}
\begin{equation}
C=\sum_{SN} \frac {1}{\sigma_i^2} \; .
\end{equation}
Eq. (\ref{eq:chi2}) has a minimum at the nuisance parameter value
$M^{\prime}=B/C$ which contains information of the values of $h$ and
$M$. That is to say, one can find the values of $h$ and $M$ when one
of them is known. However, in the literatures
\cite{ref:smallomega,ref:BDSN1,ref:BDSN2,ref:BDSN3,ref:JCAPXU,ref:SNchi2},
the expression

\begin{equation}
\chi^2_{SN}(p_s,B/C)=A-(B^2/C)\label{eq:chi2SN}
\end{equation}
is used usually in the likelihood analysis, which is up to a
constant to Eq. (\ref{eq:chi2mar}). In this case, the results will
not be affected when the distribution of $M^{\prime}$ is flat.

%%%%%%%%%%%%%%%%%%%%%%%%%%%%    BAO
\subsection{BAO}
The BAO are detected in the clustering of the combined 2dFGRS and
SDSS main galaxy samples, and measure the distance-redshift relation
at $z = 0.2$. BAO in the clustering of the SDSS luminous red
galaxies measure the distance-redshift relation at $z = 0.35$. The
observed scale of the BAO calculated from these samples and from the
combined sample are jointly analyzed using estimates of the
correlated errors, to constrain the form of the distance measure
$D_V(z)$ \cite{ref:Okumura2007,ref:Eisenstein05,ref:Percival1,ref:Percival2}
\begin{equation}
D_V(z)=\left[(1+z)^2 D^2_A(z) \frac{cz}{H(z)}\right]^{1/3},
\label{eq:DV}
\end{equation}
where $D_A(z)$ is the proper (not comoving) angular diameter
distance, which has the following relation with $d_{L}(z)$
\begin{equation}
D_A(z)=\frac{d_{L}(z)}{(1+z)^2}.
\end{equation}
Matching the BAO to have the same measured scale at all redshifts
then gives \cite{ref:Percival2}
\begin{equation}
D_{V}(0.35)/D_{V}(0.2)=1.736\pm0.065.
\end{equation}
Then, the $\chi^2_{BAO}(p_s)$ is given as
\begin{equation}
\chi^2_{BAO}(p_s)=\frac{\left[D_{V}(0.35)/D_{V}(0.2)-1.736\right]^2}{0.065^2}\label{eq:chi2BAO}.
\end{equation}

\subsection{CMB shift Parameter R}

The CMB shift parameter $R$ is given by \cite{ref:Bond1997}
\begin{equation}
R(z_{\ast})=\sqrt{\Omega_m H^2_0}(1+z_{\ast})D_A(z_{\ast})/c
\end{equation}
which is related to the second distance ratio
$D_A(z_\ast)H(z_\ast)/c$ by a factor $\sqrt{1+z_{\ast}}$. Here the
redshift $z_{\ast}$ (the decoupling epoch of photons) is obtained by
using the fitting function \cite{Hu:1995uz}
\begin{equation}
z_{\ast}=1048\left[1+0.00124(\Omega_bh^2)^{-0.738}\right]\left[1+g_1(\Omega_m
h^2)^{g_2}\right],
\end{equation}
where the functions $g_1$ and $g_2$ are given as
\begin{eqnarray}
g_1&=&0.0783(\Omega_bh^2)^{-0.238}\left(1+ 39.5(\Omega_bh^2)^{0.763}\right)^{-1}, \\
g_2&=&0.560\left(1+ 21.1(\Omega_bh^2)^{1.81}\right)^{-1}.
\end{eqnarray}
The 5-year {\it WMAP} data of $R(z_{\ast})=1.710\pm0.019$
\cite{ref:Komatsu2008} will be used as constraint from CMB, then the
$\chi^2_{CMB}(p_s)$ is given as
\begin{equation}
\chi^2_{CMB}(p_s)=\frac{(R(z_{\ast})-1.710)^2}{0.019^2}\label{eq:chi2CMB}.
\end{equation}

\end{document}